\documentclass[a4paper,fleqn,usenatbib]{mnras}

\usepackage[T1]{fontenc}
\usepackage{ae,aecompl}
\usepackage{color}
\usepackage{ulem} 

\usepackage{graphicx}	
\usepackage{amsmath}	
\usepackage{amssymb}	
\usepackage{gensymb}

\title[Thick-disc Model for NGC 247]{Thick-disc model to explain the 
spectral state transition in NGC 247}

\author[Guo et al.]{Jing Guo,$^{1}$,
Mouyuan Sun,$^{2,3}$\thanks{E-mail: ericsun@ustc.edu.cn}
Wei-Min Gu,$^{1}$\thanks{E-mail: guwm@xmu.edu.cn}
Tuan Yi,$^{1}$
\\
$^{1}$Department of Astronomy, Xiamen 
University, Xiamen, Fujian 361005, China\\
$^{2}$CAS Key Laboratory for Research in Galaxies and Cosmology, 
Department of Astronomy, \\
University of Science and Technology of China, Hefei 230026, 
China\\
$^{3}$School of Astronomy and Space Science, University of Science and 
Technology of China, Hefei 230026, China
}

\date{Accepted XXX. Received YYY; in original form ZZZ}

\pubyear{2018}

\begin{document}
\label{firstpage}
\pagerange{\pageref{firstpage}--\pageref{lastpage}}
\maketitle

\begin{abstract}
We propose the thick-disc model of \citet{Gu2016} to interpret the 
transition between soft ultraluminous state (SUL) and supersoft 
ultraluminous (SSUL) state in NGC 247. As accretion rate increases, 
the inner disc will puff up and act as shield to block the innermost 
X-ray emission regions and absorb both soft and hard 
X-ray photons. The absorbed X-ray emission will be re-radiated as 
a much softer blackbody X-ray spectrum.  Hence NGC 247 shows 
flux dips in the hard X-ray band and transits from the SUL state to the 
SSUL state. 
The $\sim 200$s transition timescale can be explained by the viscous 
timescale. According to our model, the inner disc in the super-soft 
state is thicker and has smaller viscous timescale than in the soft 
state. X-ray flux variability, which is assumed to be driven by 
accretion rate fluctuations, might be viscous time-scale invariant. 
Therefore, in the SSUL state, NGC 247 is more variable. The 
bolometric luminosity is saturated in the thick disc; the observed 
radius-temperature relation can therefore be naturally explained.
\end{abstract}


\section{Introduction} \label{sec:intro}

Typical ultraluminous X-ray sources (ULXs) are non-nuclear 
accreting X-ray sources whose luminosity exceeds the Eddington luminosity
\citep[for review, see][]{Di Stefano2003,Feng2011,Kaaret2017}. Nevertheless, 
their physical properties remain unclear since being detected for $40$ years 
\citep{Fabbiano1989}. It is likely that ULXs are powered by super-Eddington 
accretion onto stellar-mass black holes \citep{Gladstone2009}. A few ULXs 
are considered as intermediate-mass black holes candidates, like HLX-1 
\citep{Farrell2009,Godet2009,Sun2016}, or neutron stars \citep{Bachetti2014}. 
There is a special subclass of ULXs that show very soft spectra, namely, 
ultraluminous supersoft X-ray sources (ULSs) \citep{Di Stefano2004}.

Two ULX/ULS unification models, which are based on super-Eddington accretion 
onto stellar-mass black hole, are proposed \citep{Urquhart2016,Gu2016}. In 
both scenarios, the compact X-ray emission regions are obscured by high column 
density gas. In the first model, powerful outflows can block our line of sight 
\citep{Urquhart2016}. In the second model, \cite{Gu2016} argued that, for super 
Eddington ratio sources, the accretion disc is geometrically thick and acts as 
a ``shielding'' gas (a similar scenario is also proposed to explain weak-line 
quasars; \citealp[see][]{Wu2011,Luo2015}). Therefore, at certain inclination 
angles, hard X-ray photons are absorbed and thermalized; these ULXs appear as 
ULSs. Though with different mechanisms, the two models argue that ULSs tend 
to be more edge-on and have larger accretion rates compared with ULXs.

ULXs are not stationary systems. On the contrary, these sources suffer significant 
X-ray variability \citep[e.g.,][]{Kubota2001,Feng2006,Roberts2006,Kaaret2009, 
Weng2018}. It is possible that a ULX can transit to ULS and vice versa. Indeed, 
by analyzing data of \textit{XMM-Newton}, \textit{Chandra}, \textit{Swift} and 
\textit{Hubble Space Telescope} observations, \citet{Feng2016} first found the 
transitions between the supersoft ultraluminous (SSUL) regime and the soft 
ultraluminous regime (SUL) in the spectra of NGC 247. In the 
SSUL regime, the X-ray spectrum of NGC 247 can be well characterized by a 
cool (i.e., the temperature $T\la 0.1$ keV) blackbody component which is 
similar to other ULSs. For the first time, the X-ray spectrum of NGC 247 in the 
SUL regime shows a strong power-law component with flux comparable to the soft 
blackbody emission. Therefore, as pointed out by \cite{Feng2016}, 
NGC 247 might not always be classified as a ULS and they use SSUL and SUL to 
refer to this source. Its hard spectrum in the SUL regime might be similar to that 
of a ULX. 

The observational facts of the SUL/SSUL transition can be summarized as follows.
\begin{itemize}
\item[1.] as NGC 247 enters the SSUL regime, the $0.2$--$10$ keV X-ray flux decreases.
\item[2.] the transition timescale is $\sim 200$ s.
\item[3.] the blackbody temperature and the radius are anti-correlated, i.e., 
$R_{\mathrm{bb}}\propto T_{\mathrm{bb}}^{-2.8\pm0.3}$.
\item[4.] the X-ray flux is more variable in SSUL than in SUL.

\end{itemize}
\citet{Feng2016} argued that their results can be explained by the outflow model. 
In this work, we propose an alternative model which is base on \citet{Gu2016} to 
explain the observed characters in NGC 247. This paper is formatted as follows. 
In Section~\ref{sec:equa}, we introduce our thick-disc model. In Section~\ref{sec:conc}, 
we make conclusions and discussion.

\section{Thick-Disc Model} \label{sec:equa}
\begin{figure}
\includegraphics[width=\columnwidth]{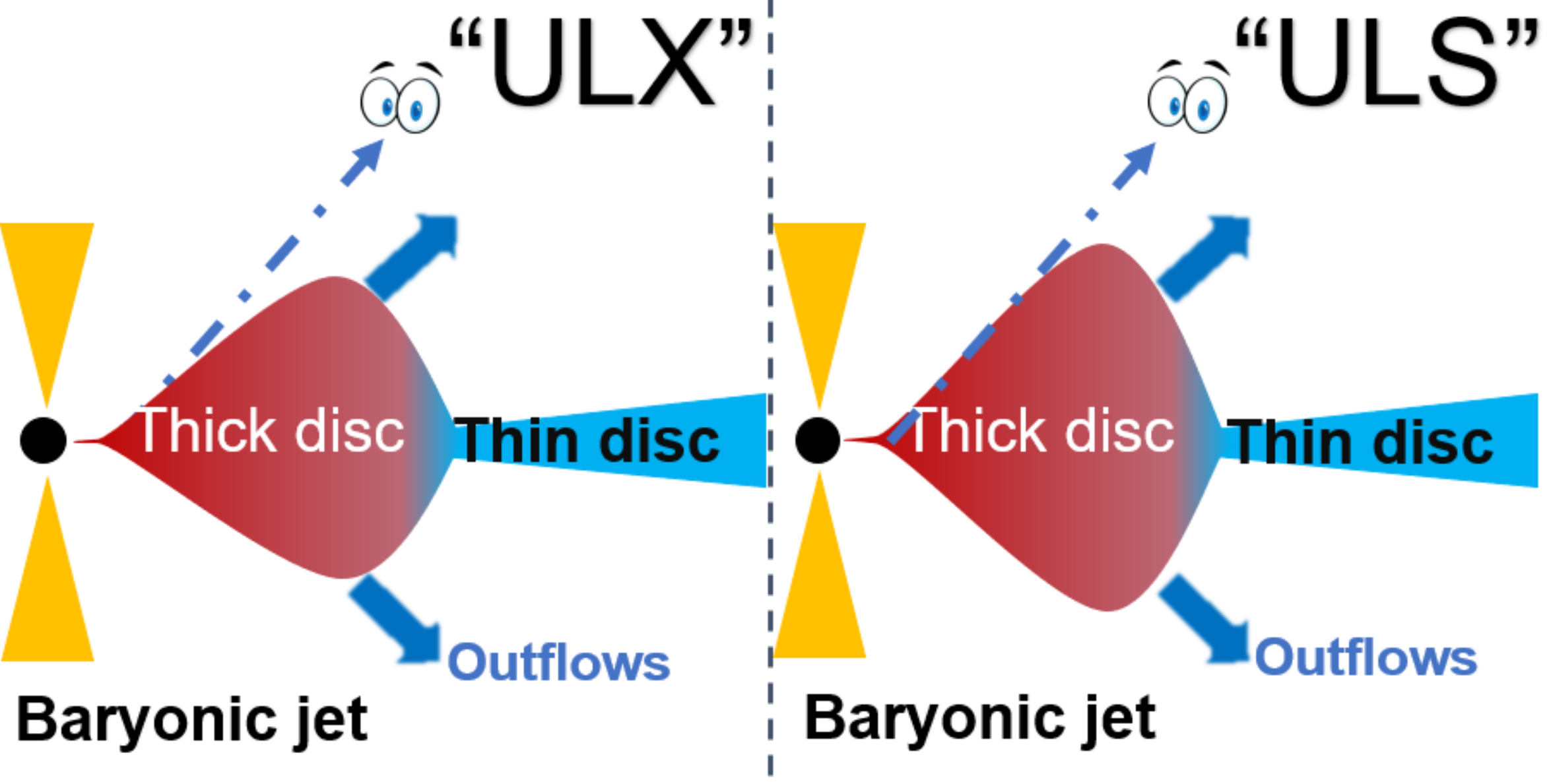}
\caption{Illustration of the thick-disc model of \citealt{Gu2016}. 
With sufficiently high accretion rate, 
the inner disc puffs up while the outer one remains geometrically thin. Significant 
outflows are also expected; the baryonic jet is launched near the BH which is 
responsible for the highly blueshifted emission lines \citep{Liu2015}. If the 
inclination angle is small (i.e., $\lesssim 25\degree$), we can directly 
observe the hard X-ray emission from the 
innermost regions; the source is identified as an ``ULX". As accretion rate increases, 
the inner disc is thicker;  the thicker disc blocks our view of the innermost X-ray 
regions. We instead only detect very soft X-ray emission produced 
from the photosphere. Hence, the source appears as an ``ULS".}
\label{F:01}
\end{figure}

We propose the thick-disc model to explain the SUL/SSUL transition (see Fig.~\ref{F:01}). 
In our model, the accretion disc can be divided into two parts. 
The inner part of the disc puffs up because of super-Eddington accretion 
\citep{Abr1988,Gu2009,Gu2012}. Such geometrical structure is also confirmed by 
recent numerical simulations \citep{Narayan2016}. Meanwhile, significant outflows 
are well expected \citep[see, e.g.,][]{Ohsuga2011,Gu2015,Sa2015}. The outer part 
(i.e., $R\gtrsim R_\mathrm{tr}$) can be described by the classical thin disc theory 
\citep{Shakura1973}. In our model, the hot innermost region could emit hard X-ray 
photons. If the inclination angle is small enough (i.e., more face-on), we can 
directly detect hard X-ray emission from the innermost regions. On the other hand, 
if the inclination is not small (i.e., more edge-on), most of hard and 
soft X-ray photons could be obscured by the thick disc. Therefore, we can only detect 
much softer X-ray photons that are produced in the effective absorption 
photosphere. The critical inclination angle $\theta_{c}$ should depend 
on the vertical structure of the thick disk. According to the numerical simulations of 
\cite{Narayan2016}, for the case of $\dot{m}=11$, $\theta_{c}\sim 25\degree$, 
which is adopted by \cite{Gu2016}. 

\begin{figure}
\includegraphics[width=\columnwidth]{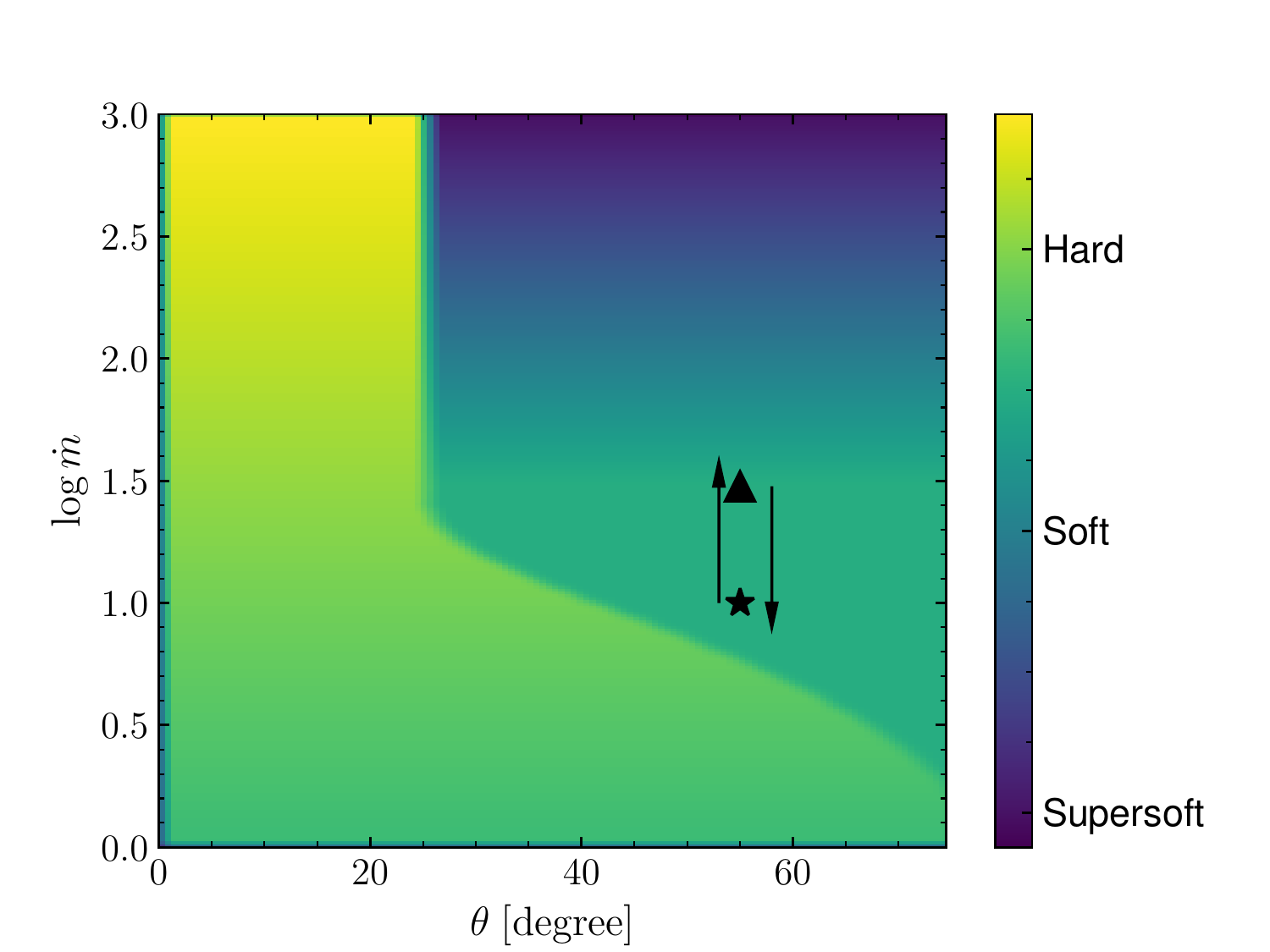}
\caption{Illustration of the unification of different spectral types of ULXs. Qualitatively, 
the star represent the SSUL regime, and the triangle represent the SUL regime. For a given 
source (e.g., NGC 247), the inclination angle ($\theta \gtrsim 25\degree$) 
is expected to be fixed. However, 
the accretion rate can vary significantly on viscous timescales. When the accretion rate 
fluctuates, the transitions between SSUL and SUL occur. }
\label{F:02}
\end{figure}

In the SUL regime (i.e., the star symbol in Fig.~\ref{F:02}) of NGC 247, our line of 
sight is assumed to be close to the edge of the thick disc (i.e., the left panel of 
Fig.~\ref{F:01}). As accretion rate increases (i.e., the triangle symbol in Fig.~\ref{F:02}), 
the inner part of the accretion disc will puff up into an even thicker one 
and block our line of sight (i.e., the right panel of Fig.~\ref{F:01}) and vice versa. 
The puffed-up disc can absorb innermost X-ray emission and re-radiate 
a much softer blackbody X-ray spectrum from its photosphere. 
Therefore, NGC 247 undergoes a transition between SUL and SSUL.

In the SSUL regime, we expect that the $0.2$--$10$ keV X-ray flux to be suppressed since 
the innermost region is blocked by the thick disc. Such suppression  increases with 
increasing energy, which is consistent with the observational fact $\#1$ \citep[see fig. 
1 of][]{Feng2016}.

The transition timescale should be the viscous timescale (see Eqs. 5.68 and 5.89 of 
\citet{Frank2002})
\begin{equation}\label{E:01}
t_\mathrm{vis} \sim 110\frac{0.01}{\alpha} \left(\frac{R}{H}\right)^2 \left(\frac{M_\mathrm{BH}}{23 
M_\odot}\right)^{-\frac{1}{2}} \left(\frac{R}{1.4\times10^{9} \ \mathrm{cm}} \right)^{\frac{3}{2}} \ 
\mathrm{s} \,
\end{equation}
where $\alpha$, $M_\mathrm{BH}$, $H$ and $R$ are the dimensionless viscosity parameter, 
the black hole mass, the height of thick disc and the transition radius, respectively. For 
geometrically thick disc, we can use $R/H \sim 1$.   

\begin{figure}
\includegraphics[width=\columnwidth]{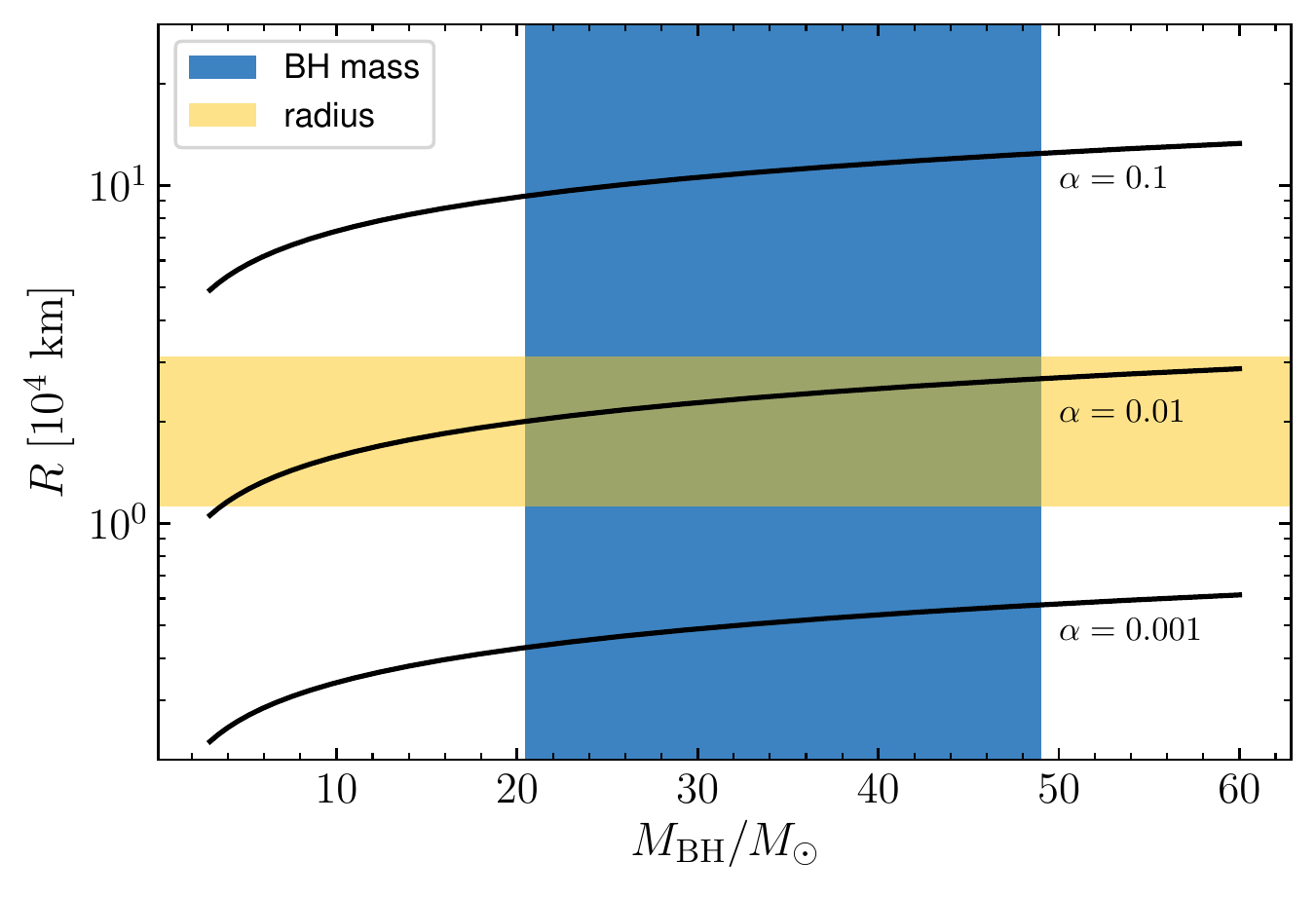}
\caption{The allowed parameter space for NGC 247. The blue shaded region is for the 
$16$th-$84$th percentiles of the ten inferred $M_{\mathrm{BH}}$(i.e., by adopting $L_{\mathrm{bol}} = 
1.7 L_{\mathrm{Edd}}$). The yellow shaded region corresponds to the $16$th-$84$th percentiles of the 
ten estimates of $R_{\mathrm{tr}}$ \citep{Feng2016}. The three solid lines indicate the relation between 
$R_{\mathrm{tr}}$ and $M_{\mathrm{BH}}$ if the transition timescale (i.e., $\sim200$ s) corresponds to 
the viscous timescale at $R_{\mathrm{tr}}$ (i.e., Eq. \ref{E:01}). }
\label{F:03}
\end{figure}

The exact value of the black hole mass of NGC 247 remains unknown. In the 
thick disc model of \citet{Gu2016}, the bolometric luminosity $L_{\mathrm{bol}}\sim 
1.7L_{\mathrm{Edd}}=2.14\times 10^{38} (M_{\mathrm{BH}}/M_{\odot})\ \mathrm{erg\ s^{-1}}$. 
We can use this relation to infer $M_{\mathrm{BH}}$. NGC 247 is a highly variable source, 
therefore the inferred $M_{\mathrm{BH}}$ changes for the ten observations. The $16$th-$84$th 
percentiles of the ten inferred $M_{\mathrm{BH}}$ are shown as the blue shaded region in 
Fig. \ref{F:03}. Meanwhile, $R_{\mathrm{tr}}$ can be constrained by fitting the ten X-ray spectra 
\citep{Feng2016}; the $16$th-$84$th percentiles of the ten estimates of $R_{\mathrm{tr}}$ are 
shown as the yellow shaded region of Fig. \ref{F:03}. The overlapping region is the allowed 
parameter space for NGC 247. 

According to our model, the viscous timescale at $R_{\mathrm{tr}}$ (i.e., Eq. \ref{E:01}) 
should be $\sim 200$ s (see the observational fact $\#2$). If so, the required 
$R_{\mathrm{tr}}$-$M_{\mathrm{BH}}$ relations for different choices of $\alpha$ are shown as 
solid lines in Fig. \ref{F:03}. It is evident that, if $\alpha\sim 0.01$, our model 
can naturally explain the observed $200$ s transition timescale. The dimensionless viscosity 
parameter $\alpha$ is determined by the magnetorotational instability turbulence and can only 
be inferred from MHD numerical simulations or observations. Observationally, $\alpha \sim 
0.1$ is estimated from outbursts of Dwarf nova and X-ray transients \citep{King2007}. However, 
MHD simulations seem prefer a much lower $\alpha \sim 0.01$ even in the radiation-pressure 
dominated regions \citep[for a summary of the value of $\alpha$ in MHD simulations, see figure 
1 of][and references therein]{Blaes2014}.

For NGC 247, the radius--temperature relation is $R_\mathrm{bb} \propto T_\mathrm{bb}^{-2.8\pm 
0.3}$. According to our thick-disc model, $R_{\mathrm{bb}}$ scales as $T_{\mathrm{bb}}^{-2}$. 
Therefore, our model and the data are consistent within the $3\sigma$ uncertainty.

X-ray variability might be caused by the fluctuation of accretion rate \citep{Lyubarskii1997}. 
If so, it is natural to expect that a viscous time-scale invariant X-ray variability. In the 
SSUL regime, the disc is thicker than that of the SUL regime. That is, the viscous timescale 
is smaller in the SSUL regime. Therefore, for fixed observational timescales, the X-ray flux 
of NGC 247 is more variable in the SSUL regime. In addition, if $\dot{m}$ 
increases to $\sim 115$, the blackbody temperature would decrease to be less than $50$ 
eV \citep[Eq. (5) of][]{Gu2016}. As a consequence, such sources become invisible to X-ray 
telescopes (e.g., \textit{Chandra}) and show extreme X-ray variability and transient behaviours.

\section{Conclusions and discussion} \label{sec:conc}

In this paper, we propose the thick-disc model of \citet{Gu2016} to interpret the observational 
facts of NGC 247. As the accretion rate fluctuates, the thickness of the puffed-up disc changes. 
As a result, the puffed-up disc can block or leave our line-of-sight. Such a model can explain 
the transition between SUL and SSUL regimes in NGC 247.

As pointed out by \cite{Feng2016}, the outflow model can also explain some 
observational aspects of NGC 247, including the $R_{\mathrm{bb}}$-$T_{\mathrm{bb}}$ relation. 
The $\sim 200$s timescale and variability might be driven by some instabilities and/or turbulence 
in the outflow; alternatively, the disc instabilities we mentioned in Section~\ref{sec:equa} might 
also induce some variations of outflow properties. So far, there is no solid evidence for or against 
the outflow model or our thick-disc model.  

Compared with the outflow model, the required accretion rate in our model is much lower. To 
illustrate this fact, we estimate the accretion rate via Eq. (5) of \citet{Gu2016} and find 
that $\dot{m}\sim 10$--$30$, which is one order of magnitude lower than that of the outflow 
model \citep[for NGC 247, see fig.~5 of][]{Feng2016}. 

With such high accretion rates, powerful outflows are inevitable both from theoretical considerations 
\citep[e.g.,][]{Weng2011,Gu2015} and numerical simulations \cite[e.g.,][]{Ohsuga2011,Sa2015}.
Meanwhile, blueshifted emission lines are also observed in some ULXs \citep{Pinto2016,Walton2016}. 
In principle, such outflows might also act as shield and increase our line-of-sight optical depth. It is 
quite possible that both the thick disc and the outflows play significant roles in obscuring the inner 
X-ray emission regions (see Fig.~\ref{F:01}). If so, the critical accretion rate for the SSUL regime
could be even lower \citep{Gu2016}. 

In future, our thick-disc model should be observationally distinguishable from 
the outflow one since the required mass rate are quite different in the two models. We propose 
a few possible observational tests. First, the two models can be distinguished if we can measure 
the mass outflow rate. Both the two models predict massive outflows; however, the mass 
outflow rate of our thick-disc model is much lower than that of the outflow model. Outflow 
signatures, i.e., absorption/emission lines, are indeed detected in NGC 247, several other 
ULSs \citep{Urquhart2016} and ULXs \citep{Pinto2016, Walton2016, Kosec2018}. 
According to our model, the lower mass rate outflows are responsible for the absorption 
features (around $1$ keV). If the mass rate of such outflow can be inferred from these features, 
our model and the outflow model can be distinguished. The low 
signal-to-noise data prevent us from constraining the mass outflow rate. Future facilities, 
e.g., \textit{HUBs}\footnote{For details, please refer to 
\url{http://heat.tsinghua.edu.cn/~hubs/en/index.html}} and \textit{Athena} \citep{Nandra2013}, 
can provide high-resolution soft X-ray spectra and might be able to measure the mass outflow 
rate. In addition, huge bubbles are observed around some ULXs \citep[e.g.,][]{Kaaret2004, 
Kaaret_C2009, Moon2011, Cseh2012}; if such bubbles are inflated by outflows, we might also 
be able to estimate the mass outflow rate by considering their ages, kinematics and total 
energies \citep[e.g.,][]{Pakull2006, Siwek2017}. However, similar bubbles are not observed for ULSs. 
Future possible discoveries of ULS bubbles would be very interesting and can verify the ULS 
models. The two models can also be tested by the intrinsic fraction of ULSs to ULXs. 
According to both models, ULSs tend to have larger accretion rate and higher inclination angle 
with respect to ULXs. The accretion rate distribution for ULXs can be inferred from binary 
population synthesis models \citep[e.g.,][]{Madhusudhan2008, Pavl2017}; the distribution drops 
rapidly at high accretion rate end. If so, our thick-disc model predicts a higher fraction of ULSs 
to ULXs than that of the outflow model because the required accretion rate to be an ULS in 
our model is roughly one order of magnitude lower. If we adopt the accretion 
rate distribution of figure 10 of \cite{Madhusudhan2008} for stellar black holes (by assuming 
an average $M_{\mathrm{BH}}=10M_{\odot}$ and the radiative efficiency of $0.1$), the ratio of the fraction of 
systems with $\dot{m}>30$ to those with $30>\dot{m}>1$ is $1/32$. If we take this ratio and the 
fact that the viewing angle of an ULS must be larger than $25$ degree, the expected ratio of 
number of ULS to that of ULX is about $1/45$. This ratio is roughly consistent with the 
observed one, albeit with large uncertainties in both theoretical and observational (because the 
sample size of ULS is quite small) ones. 
All in all, future observations of ULSs can decipher the super-Eddington accretion physics.

\section*{Acknowledgements}
We thank the anonymous referee for his/her helpful comments that improved the paper.
We are grateful to Hua Feng, Xiangdong Li, Dabin Lin and Yaping Li for valuable discussion. This 
work was supported by the National Natural Science Foundation of China under grants 11573023, 
11333004 and 11603022. M.Y.S. acknowledges the support from the China Postdoctoral Science 
Foundation (2016M600485).

\end{document}